\documentclass[multphys,vecphys]{svmult}

\usepackage{makeidx}         
\usepackage{graphicx}        
\usepackage{multicol}        
\usepackage[bottom]{footmisc}

\makeindex             

\begin{document}

\title*{Kinematics of Globular Cluster Systems}
\author{Aaron J. Romanowsky
}
\institute{Departamento de F\'{i}sica, Universidad de Concepci\'{o}n, Casilla 160-C, Concepci\'{o}n, Chile
{\hskip 0.1cm}
\texttt{romanow@astro-udec.cl}
}
%
%
\maketitle

I review the field of globular cluster system (GCS) kinematics,
including a brief primer on observational methods.
The kinematical structures of spiral galaxy GCSs 
so far appear to be broadly similar.
The inferred rotation and mass profiles of elliptical galaxy halos exhibit
a diversity of behaviors, 
requiring more systematic observational and theoretical studies.

\section{Introduction}
\label{sec:intro}

Globular clusters (GCs) are well known as useful probes of the formation histories
of galaxies, especially at early times.
{\it Kinematics} of globular cluster systems (GCSs) brings an extra dimension to these probes,
helping firstly to {\it distinguish GC subsystems} by their distinctive kinematical signatures.
The need for such information is highlighted by 
recent doubts about bimodalities in GC metallicity distributions.
Kinematical parameters can also help identify
{\it correlations between GCs and field stars} within galaxies, and
so determine which objects had a common origin.
The present-day motions of GCs can provide {\it signatures of their formation
and evolution,}
as different physical processes will lead to different orbital properties.
Last but not least, the residence of GCs at very large radii around galaxies
makes them excellent {\it halo mass tracers}.

The informative capacity of GCS kinematics can be illustrated by the well-known case 
of the Milky Way, where
there are at least two subpopulations (the bulge/disk and halo
GCs).
The cornerstone characteristics of these subpopulations are their
spatial distributions, metallicities, and kinematics 
(e.g. \cite{romanow:zinncote,romanow:pritzl05}): 
without all three clues, the distinction would be much less clear.
One may make further decompositions into ``thin/thick'' disk GCs, 
and ``young/old'' halo GCs---which is a level of detail difficult to
achieve in external galaxies.
Note that these nomenclatures are not arbitrary, but
connect to the picture that GC and field star formation are associated.
GCs have also proved valuable for measuring the mass of the Milky Way halo
(e.g. \cite{romanow:wilkinson99,romanow:sakamoto03}).


\section{GCS kinematics observations}
\label{sec:obs}

The study of GCS kinematics beyond the Local Group began in the 1980s with
the use of 4-m-class telescopes, providing dozens of velocities in
the bright Virgo ellipticals M87 and M49
\cite{romanow:mould,romanow:huchra87},
and expanded in the 1990s with the use of 8-m-class telescopes to obtain hundreds of
velocities out to distances of $\sim$~20 Mpc
\cite{romanow:cohen,romanow:cote03,romanow:richtler04}.
Today, the sample of well-studied galaxies is still small,
so it is worth reviewing the observational challenges.

\subsection{Observational requirements and issues}

The ideal starting point for GCS kinematics observations is
{\it wide-field imaging}, allowing candidate GCs to be identified
far out into a galaxy's halo.
Imaging that reaches beyond the extent of the GCS can further
be used to measure the ``contamination'' from stars and galaxies.
Although relatively shallow images are enough to find the bright GCs
which are accessible to spectroscopy,
it is important to obtain much deeper images, since GCS dynamical models
need a well-known spatial distribution.
There have so far not been many GCS imaging studies published with
the $\sim$0.5$^{\circ}$ field of view needed for nearby galaxies
\cite{romanow:dirschrhode}.

Contamination levels increase rapidly as one moves to studying outer
halo GCs, so any available countermeasure should be employed.
{\it High spatial resolution} imaging is profitable but not usually obtainable.
Multiple ACS pointings are one avenue to a reasonably wide field-of-view around
a galaxy.
From the ground, good seeing can permit GCs to be resolved out
to 5~Mpc 
(e.g. \cite{romanow:gomez06}).

The standard tool for GC identification is {\it color selection}.
The Washington $C$ filter is especially powerful in identifying
unresolved faint blue galaxies by their UV excess, while
three-band photometry helps even more with object discrimination,
and broader color baselines are better at resolving
GC bimodality:
$(C-R)$ has three times the resolution of $(V-I)$
\cite{romanow:dirschrhode}.

For acquiring the GC velocities, efficient wide-field
multi-plexed spectrographs are needed.
For astrometric reasons it is ideal to
have {\it pre-images} of the GCs taken on the same instrument,
which entails an imaging spectrograph.
However, many observatories cannot provide pre-imaging
without introducing a year's delay in the spectroscopy.
Another observational hurdle is {\it competition}:
acquiring hundreds of GC velocities involves
$\sim$~10 hours of 8-m dark time,
often in an especially oversubscribed period
(R.A. $\sim$ 12h).

A key issue in GC spectroscopy is {\it sky subtraction},
which is normally more accurate with slits than with fibers.
It is not yet clear that the benefits of nod-and-shuffle 
outweigh the complications.
A stable high-resolution fiber spectrograph such as UT+FLAMES/GIRAFFE
can provide velocities as accurate as 5~km~s$^{-1}$, which is useful
for galaxies with very low mass or low rotation.
The wavelength range used
is typically $\sim$~4000--6000\AA{},
but an option for brighter skies is the
Ca~II triplet at $\sim$~8600\AA{}
(e.g. \cite{romanow:goudfrooij01}).
Well-matched {\it template spectra} are essential to finding accurate velocities.
Synthetic spectra or observations of Galactic stars can be used,
but better matches come from Local Group GCs, or early-type dwarf galaxies.
A final impediment to progress is the rarity of
fully-operational {\it data-reduction pipelines} for
multi-object spectrographs.

\begin{figure}[t!]
\centering
\includegraphics[height=5.85cm]{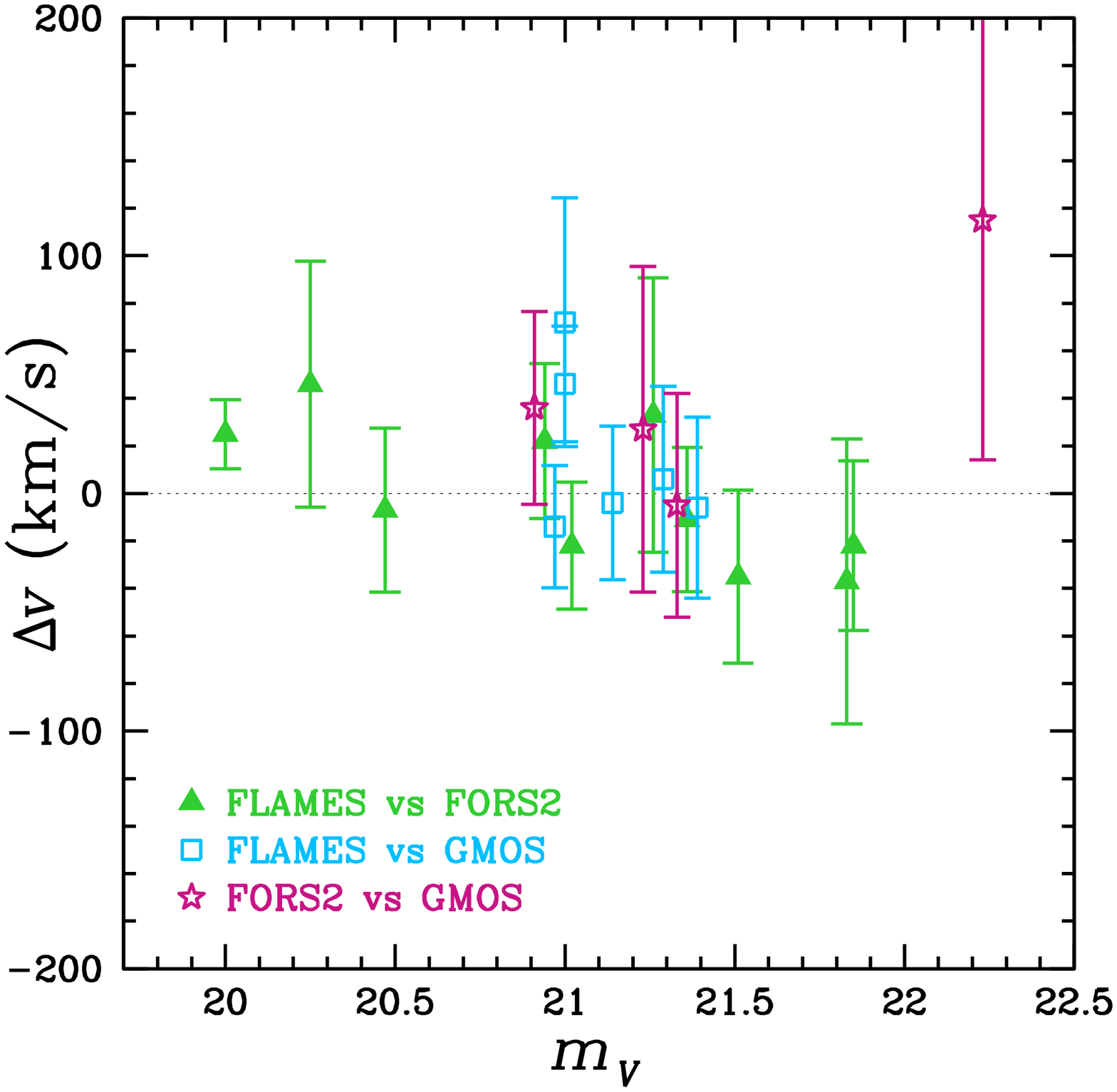}
\includegraphics[height=5.85cm]{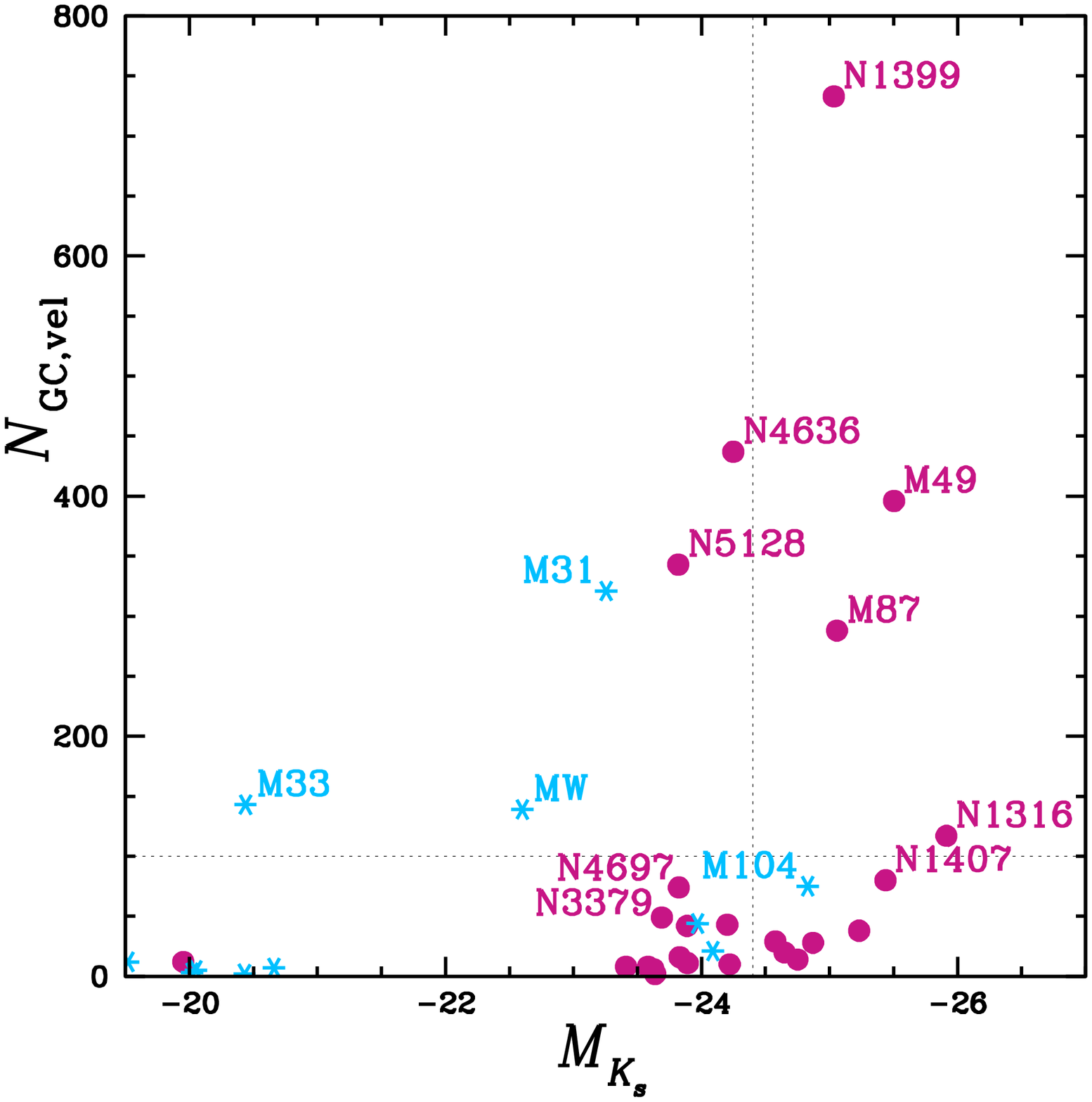}
\caption{
{\it Left:}
GC velocity comparisons around the galaxy NGC~3379, as a function of magnitude,
from different instruments 
\cite{romanow:puzia04,romanow:piercebergond}.
{\it Right:}
Number of GC velocities acquired as a function of galaxy magnitude.
Different colors and symbols show early-type and late-type galaxies.
A ``critical'' value of 100 GCs is shown by a horizontal line, and
a ``typical'' early-type galaxy luminosity $L^*$ is shown by a vertical line.
}
\label{romanow:fig:1}       
\end{figure}

With improvements in telescopes, instruments, and techniques, the observational 
productivity and accuracy of GC kinematics
have increased dramatically in recent years.
E.g. at GC magnitudes of $R\sim 20$, the velocity uncertainties have improved from
$\sim$~100~km~s$^{-1}$ to
$\sim$~20~km~s$^{-1}$ \cite{romanow:richtler04}.
The measurement reliability
is shown by the consistency
within the stated errors of the {\it absolute} velocities (no systematic shift applied)
when comparing 8-m telescope observations of GCs around NGC~3379 down to $V=22$
(see Fig.~\ref{romanow:fig:1}, left).
The most distant galaxy yet reached with GCS spectroscopy is NGC~3311
at 50~Mpc, using UT3+VIMOS
(T. Richtler et al, in prep).

To compare GC and stellar dynamics in a galaxy, 
one needs {\it observations of the stellar kinematics}.
In galaxy halos outside the Local Group,
such observations are currently feasible only by using
{\it planetary nebulae}
(PNe) as stellar proxies---an approach which has taken off in recent years
(e.g. \cite{romanow:mendez01,romanow:romanowsky03}).

\subsection{Observational studies}

To credibly measure rotation without assuming a preferred axis,
or to estimate mass without assuming some orbital anisotropy,
requires $\sim$100 GC velocities.
Many GC spectroscopic studies using large telescopes
have focused on line-strengths, producing 
samples not particularly suitable
for dynamical studies.
There are now only $\sim$10 galaxies with large GC velocity samples
(see Fig.~\ref{romanow:fig:1}, right).
%
The first distant GCS studied in kinematical detail belongs to the Virgo Cluster central elliptical
M87, with 288 velocities acquired out to galactocentric distances of 45 kpc---mostly
by Keck+LRIS and CFHT+MOS
\cite{romanow:cohen,romanow:hanes}.
The {\it brightest} Virgo elliptical, M49 (=NGC~4472), 
now surpasses it with 396 velocities to 90 kpc, mainly from 
LRIS and FLAMES
(\cite{romanow:cote03}; G. Bergond et al, in prep).
%
The central Fornax Cluster elliptical, NGC~1399,
has the largest data set of any galaxy,
with $\sim$700 velocities to 90 kpc,
mainly from UT2+FORS2
\cite{romanow:richtler04}.
NGC~1407, the central Eridanus A Group elliptical,
has $\sim$100 velocities from
LRIS, FLAMES, and Clay+LDSS-3 
(P. S\'{a}nchez-Bl\'{a}zquez et al, in prep).

These ``high-end'' ellipticals have been so extensively studied because of their
very rich GCSs, but results for more ordinary ellipticals (with $\sim L^*$
luminosity) have been slow in coming.
Even 8-m-class telescopes have trouble getting large kinematic data sets
in some of these galaxies because of their limited GC populations.
An example is NGC~3379, which is a very interesting galaxy
but unfortunately has a sparse GCS.
Several observing campaigns have achieved only 49 GC velocities
\cite{romanow:puzia04,romanow:piercebergond}, although
another recent FLAMES study should double or triple this number 
(G. Bergond et al, in prep).

The first large GC kinematics study of an $L^*$ elliptical involves 
NGC~4636, at the Virgo 
outskirts.
Here UT2+FORS2 has been used to obtain 437 GC velocities out to 45 kpc
(\cite{romanow:schuberth06}; Y. Schuberth, this vol).
Further $L^*$ elliptical studies are now underway with UT2+FORS2,
Gemini+GMOS, and elsewhere.
Another extensively studied GCS is from the nearby (4 Mpc) disturbed
early-type galaxy NGC~5128.
Its ever-growing tally using 4-m telescopes is 343 GC velocities to 40 kpc
(e.g. \cite{romanow:woodley06}).

{\it Spiral} galaxies have received less attention than ellipticals.
By far the largest study is of the M31 GCS, resulting in
321 velocities, largely from the MMT and the WHT
\cite{romanow:m31}.
M33 has 143 velocities, mostly from 
WIYN
\cite{romanow:chandar02}.

\section{GCS kinematics in spiral galaxies}
\label{sec:spiral}

The available 
data in spiral galaxies permit a brief summary,
going from latest type to earliest.
In M33, there is a rotating component of young GCs,
and a weakly-rotating old GCS 
with possible
halo and disk sub-components
\cite{romanow:chandar02}.
In the Milky Way, there is a rotating 
GC component
associated with the stellar bar or bulge, and
a hot, non-rotating halo GCS.
There are ``old'' and ``young'' halo GCs with differing
kinematics \cite{romanow:dinescu99,romanow:mackey04},
and the halo {\it stars} have a velocity dispersion 
decreasing with radius,
similarly to the halo GCS
\cite{romanow:battaglia05}.
In M31, there is rotation in both the metal-poor and metal-rich GCs,
a thin-disk GCS whose age is controversial,
and again a metal-poor stellar halo with a decreasing 
dispersion profile
\cite{romanow:m31,romanow:m31b}.
In M81, there is a rotating metal-rich GC component and a non-rotating
metal-poor component
\cite{romanow:schroder02}.
In M104, there appears to be little rotation regardless of metallicity, 
although the GC sample is still small
(e.g. \cite{romanow:bridges97}).
Thus we see that a GCS kinematical structure like the Milky Way's may be
common, but perhaps not universal,
and also that there is more work to be done comparing the behavior
of halo stars and GCs.

\section{GCS rotation in elliptical galaxies}
\label{sec:rot}

While the luminous bodies of ellipticals are observed to have strikingly lower angular momenta than spirals,
the ``missing'' spin might reside in their halos.
This scenario is supported by simulations of elliptical formation via galaxy mergers,
which predict high rotation ($v/\sigma \sim 1$) outside $\sim 2 R_{\rm eff}$
(e.g. \cite{romanow:bekki05})---with the caveat that
dissipative 
baryonic effects are not yet well-studied.
A galaxy's GCS would not directly contain much of its angular momentum,
but it could be used to infer the rotation of the halo stars or dark matter.

So far there is no obvious pattern to GCS rotational properties,
except that dwarf ellipticals may be more rotationally dominated than giants
\cite{romanow:beasley06}. 
There is no consistent trend of GC rotation with metallicity,
nor a clear correlation between GC and stellar halo
rotation fields (as inferred by PN kinematics).
M87 and M49 may have large GCS kinematical twists,
as evidence perhaps of GC accretion
\cite{romanow:cote03,romanow:hanes}.
Further progress should come with more data,
placed in the context of global dynamical models and more detailed theoretical predictions,
with attention to the connections to galaxy sub-type and environment.

\section{GCS dynamical modeling}
\label{sec:dyn}

The simplest ingredient for a GCS dynamical model
is the projected velocity dispersion profile $\sigma_{\rm p}(R)$.
Inferences from $\sigma_{\rm p}$ are tricky
due to the classic mass-anisotropy degeneracy, which
can be lifted by attention to the {\it shape} of
the line-of-sight velocity distribution (LOSVD).
This requires $\sim$1000 GC velocities even assuming spherical symmetry
\cite{romanow:merritt93},
but fortunately there are additional constraints that drastically reduce the uncertainties.
The GCS surface density
may be inferred from many more GCs than have measured velocities,
and the central galaxy mass can be well-determined from stellar kinematics.
One of the key potential advantages of GCS dynamics is 
less complicated geometry than stellar dynamics.
In particular, the metal-poor GCs may lie in near-spherical distributions---although
this supposition has had little empirical testing.

\subsection{Elliptical galaxy halo masses}

The GCS $\sigma_{\rm p}(R)$ profiles of
M87, M49, NGC~1399, and NGC~1407 are all 
constant or rising with the radius.
Such behavior is clear evidence for massive dark matter halos,
or else fairly pathological anisotropy would be required.
For M87, ``orbit modeling'' has been carried out
which includes non-parametric anisotropy profiles and
LOSVD shape-fitting for the GCs and the stars
\cite{romanow:romanowsky01}.
A dark halo is found which appears to belong to the Virgo Cluster
core, with a profile in encouragingly good agreement with 
independent X-ray-based mass results
\cite{romanow:matsushita02}.
Another non-parametric study has found similar results
\cite{romanow:wu06}.
In NGC~1399, a rising GCS $\sigma_{\rm p}(R)$ may also
trace the Fornax Cluster core, 
or it may be a transient feature caused by galaxy interactions
\cite{romanow:napolitano02,romanow:bekki03}.

NGC~4636 is a typical $L^*$ elliptical, except that its
rich GCS and high X-ray luminosity suggest that it
is at the center of a group halo.
Its GCS 
kinematics imply
a group-scale dark matter halo---but less
massive than inferred from the (disturbed) X-ray-emitting gas
\cite{romanow:schuberth06}.
Studies of PN kinematics in ordinary ellipticals suggest that
such galaxies do not have the centrally-concentrated massive 
dark matter halos expected from $\Lambda$CDM
\cite{romanow:mendez01,romanow:romanowsky03,romanow:napolitano05}.
Other interpretations have been supplied
\cite{romanow:dekel05},
and it would be invaluable to have independent GCS constraints on
the same galaxies studied using PNe.

\begin{figure}[t!]
\centering
\includegraphics[height=5.85cm]{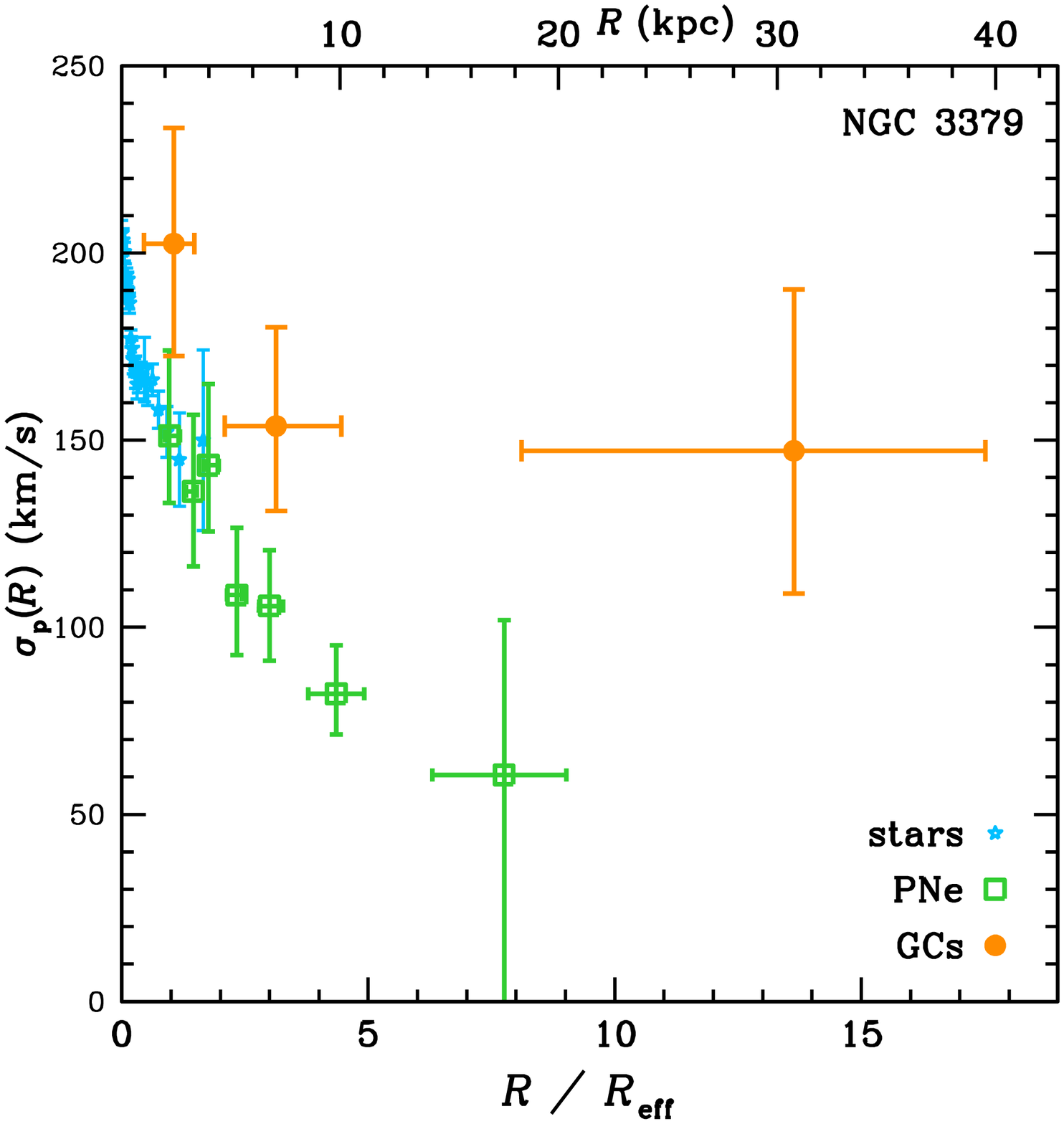}
\includegraphics[height=5.85cm]{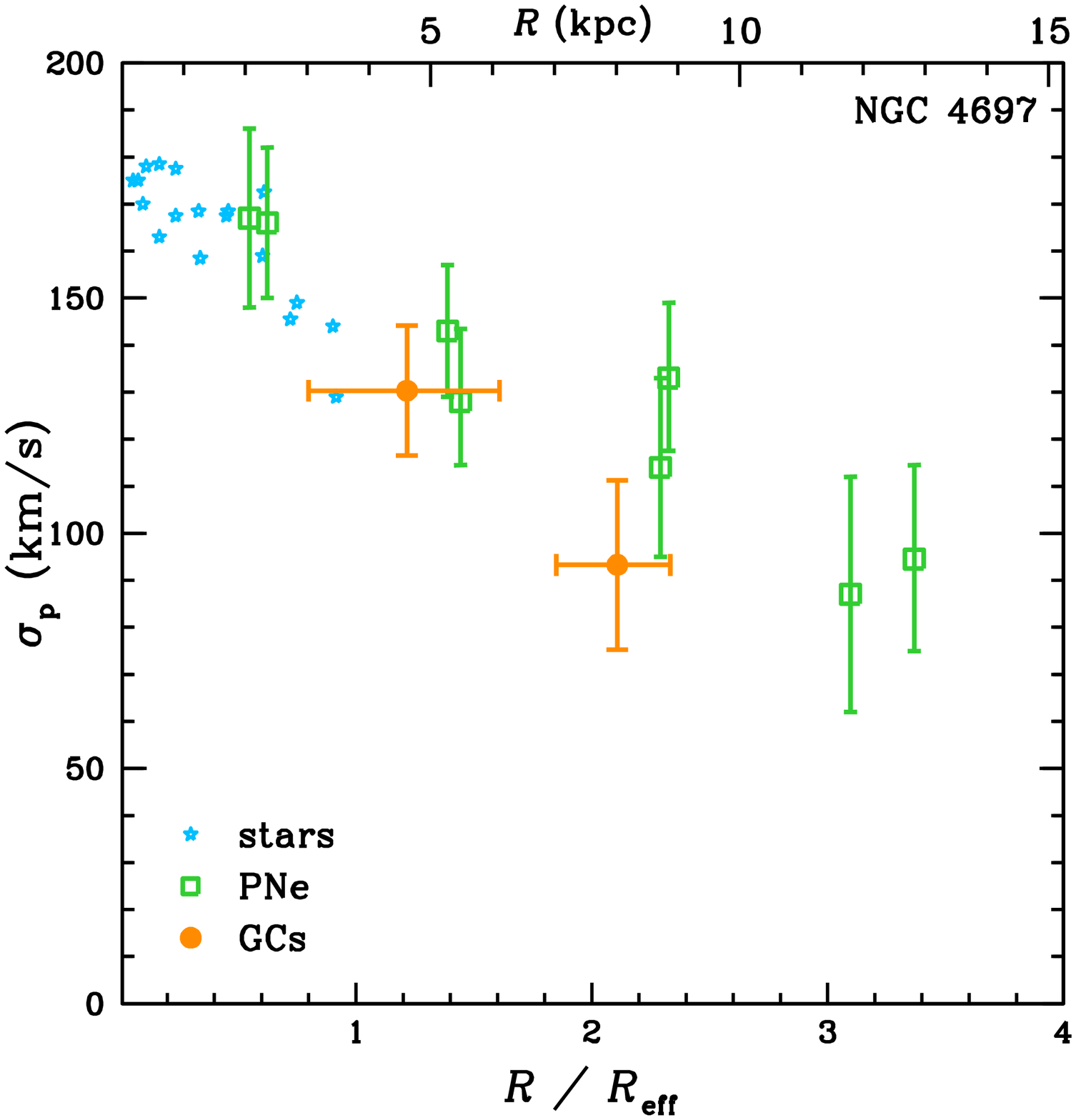}
\caption{
Projected velocity dispersion profile data in nearby $L^*$ ellipticals, for stars, 
PNe, and GCs.
{\it Left:}  
The round galaxy NGC~3379 
(\cite{romanow:puzia04,romanow:piercebergond,romanow:romanowsky03}; N. Douglas et al, in prep).
{\it Right:}
The flattened galaxy NGC~4697
(\cite{romanow:mendez01}; A. Romanowsky et al, in prep).
}
\label{romanow:fig:2}       
\end{figure}

For the poster-child low-dark-matter galaxy, NGC~3379,
there are not yet many GC velocities, but they do reach to
much larger radii than the PNe.
The GCS shows a much shallower decline in $\sigma_{\rm p}(R)$
than the PNe (see Fig.~\ref{romanow:fig:2}, left).
The difference seems to be largely due to very different
spatial and anisotropy distributions---e.g.,
GCs and PNe at the {\it same projected radius}
are sampling very different regions of the 3D potential.
The implied mass profiles from stellar, PN, and GCS dynamics are
consistent within the still-considerable uncertainties.
Intriguingly, this system also includes an HI gas ring
\cite{romanow:schneider85} whose kinematics
imply a low halo mass that is {\it not} consistent with the GCS results.
For another low-dark matter galaxy, NGC~4697,
preliminary results from Gemini+GMOS indicate 
that the GCS $\sigma_{\rm p}(R)$
{\it does} track the decline of the PNe (see Fig.~\ref{romanow:fig:2}, right).
More data at larger radii are still needed from this ongoing program.

Further results from PN and GCS studies are awaited,
but it is already clear that there is {\it great diversity} in the mass profiles
of ellipticals---which
may be contrasted with the tight Tully-Fisher relations of spirals.
Are all the obvious dark matter halos in galaxies such as M49
examples of a group or cluster halo,
while the individual halos of free-floating galaxies such as NGC~4697
are much scantier?
Or is there a large amount of scatter in halo properties relating to the
galaxies' collapse, merger, and interaction histories, coupled to
normal scaling relations?
Large, systematic dynamical surveys, as well as more progress with
simulations, are needed to address such questions.

\subsection{GCS orbital properties}

The orbits of GCs can inform us about their
formational histories, and their connections with their
host galaxy's field star populations.
To determine these properties, it is very helpful to
have independent constraints on galaxy mass profiles
(e.g. from X-ray emission) 
so that the GC kinematics can be directed
to determining the anisotropy.
Examples are M87 and M49, where
the GCs turn out to have overall fairly isotropic orbit distributions
\cite{romanow:cote03,romanow:hanes,romanow:romanowsky01}.
In M87, the metal-poor GCs appear to have slightly
tangential orbits inside $3 R_{\rm eff}$, and
slightly radial orbits in the outer parts,
while the metal-rich GCs have slightly radial orbits
everywhere.
These orbital properties imply that dynamical processes could not have
dramatically changed
the GC mass function \cite{romanow:vesperini03}.

More detailed analyses of large velocity data sets could
cross-correlate GC and stellar kinematics,
to test the notion that metal-rich GCs are coeval with
their galaxy's old metal-rich field stars.
One may also look for bimodality in dynamical phase-space that correlates
with color to support the existence of metallicity bimodality.
Further inferences about galaxy/GCS formation from GC orbital properties are in 
dire need of firmer theoretical predictions
(see e.g. \cite{romanow:bekki05} for pioneering work).
Some $\Lambda$CDM-based simulations do predict strong {\it radial anisotropy of
the stars} in galaxy halos \cite{romanow:diemand05,romanow:abadi06},
which may not accord with the {\it observed isotropy of GCSs}.
Orbit analyses can also find remnant substructures from accretion and
interaction events, as may
have already been identified in M31 and M49
\cite{romanow:cote03,romanow:perrett03}.

\subsection{Modified Newtonian Dynamics}

An alternative to dark matter for explaining mass discrepancies is
Modified Newtonian Dynamics (MOND), a theory with
much success in modeling late-type galaxies.
While so far rarely applied to ellipticals, MOND seems to be
consistent with the cases of declining PN $\sigma_{\rm p}(R)$ profiles
\cite{romanow:milgrom03}.
The greatest observational challenge to MOND may be 
the high apparent dark matter content in galaxy cluster cores.
These cores are typically inhabited by bright ellipticals with rich GCSs
whose dynamics can be used to test MOND---although the conclusions may
not be as clear-cut as hoped (\cite{romanow:schuberth06}; T. Richtler
et al, this vol).

\section{Summary}
\label{sec:sum}

GCS kinematics studies are now beginning to provide large data sets for
a wide variety of galaxies.
Such data are useful for identifying true GCS subcomponents.
In elliptical galaxies, diverse GCS rotation profiles are seen, as well as typically
constant dispersion profiles implying massive dark matter halos.
However, these results come from massive group-central ellipticals,
and new observations are addressing more ordinary ellipticals---in particular, to
see how their mass distributions square with theoretical predictions.

\printindex
\end{document}